\documentclass[aps,
superscriptaddress,showpacs,showkeys,amsmath,amssymb,floatfix]{revtex4}
\usepackage{hyperref}
\usepackage{color}
\usepackage{graphicx}
\usepackage{epsfig}
\usepackage{dcolumn}
\usepackage{bm}
\usepackage{mathtools}
\usepackage{amssymb}
\usepackage{bmpsize}
\usepackage{calrsfs}
\DeclareMathAlphabet{\pazocal}{OMS}{zplm}{m}{n}
\usepackage{amsmath}
\usepackage{subfigure}
\newcommand{\be}{\begin{equation}}
\newcommand{\ee}{\end{equation}}
\newcommand{\bea}{\begin{eqnarray}}
\newcommand{\eea}{\end{eqnarray}}
\newcommand{\bpsi}{{\bar\psi}}
\newcommand{\cL}{{\cal{L}}}

\newcommand{\hPsi}{{\hat \Psi}}
\newcommand{\hu}{{\hat u}}
\newcommand{\hv}{{\hat v}}

\newcommand{\hc}{{\hat c}}

\newcommand{\tpsi}{{\tilde \psi}}
\newcommand{\tC}{{\tilde C}}

\newcommand{\tL}{{\tilde L}}

\newcommand{\vA}{\vec A}

\newcommand{\pro}{\partial}

\newcommand{\bfr}{{\mathbf r}}

\newcommand{\bfw}{{\mathbf w}}

\newcommand{\ba}{\begin{array}}
\newcommand{\ea}{\end{array}}

\newcommand{\nn}{\nonumber}

\newcommand{\bbE}{\mathbb E}
\newcommand{\bbR}{\mathbb R}

\begin{document}
\title{Non-perturbative on-shell multiplet structure of $SU(N)$ Yang-Mills fields}
\author{Dmitriy G. Pak}
\email{drgn2@yandex.ru}
\affiliation{School of Physics and Astronomy,
 Sun Yat-sen University, Zhuhai, 519000, China}
\affiliation{Bogolyubov Laboratory of Theoretical Physics, JINR, 141980 Dubna, Russia}
\affiliation{Physical-Technical Institute of Uzbek Academy of Sciences, Tashkent 100084, Uzbekistan}
\author{Takuya Tsukioka}
\email{tsukioka@bukkyo-u.ac.jp}
\affiliation{School of Education, Bukkyo University, Kyoto 603-8301, Japan}
\author{Pengming Zhang}
\email{zhangpm5@mail.sysu.edu.cn}
\affiliation{School of Physics and Astronomy,
 Sun Yat-sen University, Zhuhai, 519000, China}
\begin{abstract}
      Color multiplets of the gauge fields and fermions on mass shell in $SU(N)$ Yang-Mills theory
are classified according to representations of the Weyl group $W(SU(N))$. The multiplet structure
of quark and gluon multiplets has been studied in the framework of a non-perturbative approach
by considering complete exact equations of motion of the $SU(N)$ Yang-Mills theory with matter
fields. An important case of singlet non-Abelian gluon solutions corresponding to one-dimensional
singlet representations of the Weyl group is revised on a rigorous mathematical basis.
We demonstrate that Weyl group as a finite color subgroup of $SU(N)$ reveals
an inherent color symmetry of quarks and gluons on mass shell which determines a universal color
multiplet structure of quark-gluon solutions in a pure $SU(N)$ Yang-Mills theory and in Abelian
projected Yang-Mills theories with quarks. The obtained results allow to introduce strict concepts
of fundamental particles, quarks and gluons, which differ drastically from the particle definitions
in the conventional perturbative Yang-Mills theory. Possible applications of our results in
non-perturbative quantum chromodynamics and hadron physics are discussed.
\end{abstract}
\maketitle
\section{Introduction}

The gauge equivalence principle is a fundamental guiding principle which allows to construct
dynamical theories of fundamental  interactions on a strict mathematical basis \cite{Weyl1929}.
A quantum chromodynamics (QCD) is a well established theory of strong forces which represents a
quantum theory of the classical gauge Yang-Mills theory with a structural color group $SU(3)$
\cite{Yang-Mills}. The conventional theory of QCD is formulated within the framework of perturbation
theory, which works well in high energy range, but has serious limitations in the infrared region. 
Due to this, several powerful non-perturbative methods have been developed based on
the lattice QCD, instanton QCD vacuum, effective chiral Lagrangians, and other approaches including 
sophisticated mathematical ideas of Ads/CFT correspondence and supersymmetry (see a short review \cite{digiac}
 and refs. therein).
\par   
So far, a strict non-perturbative formulation of QCD still represents one of
unresolved Millennium problems announced by Clay Institute of Mathematics \cite{Mill2000}.
Difficulties in constructing a rigorous quantum Yang-Mills theory arise at the very first steps of
introducing the concepts of fundamental particles, quarks and gluons. 
Besides, the QCD vacuum has been representing another long-standing problem since 1980
when Nielsen and Olesen found the quantum instability of the vacuum gluon condensate made of constant
color magnetic field \cite{N-O}. Note, that various vacuum gluon solutions like color magnetic vortices, monopoles, 
dyons etc, do not possess local stability under the quantum vacuum fluctuations. The vacuum instability problem 
has been resolved  in  \cite{plb2018} where it has been demonstrated  that microscopic structure of a
stable vacuum gluon condensate is provided by the stationary Abelian gluon solutions given in terms of 
vector spherical harmonics. Such vector harmonics describe the microscopic structure of the vacuum gluon condensate
with oscillating function of the gluon condensate density \cite{plb2023}. This result is in  
agreement with the result obtained earlier in the framework of the instanton approach to QCD \cite{dorokhov1997}. 
 Note, that recently an alternative quantum stable QCD vacuum gluon configuration 
has been proposed  in \cite{savv2}. 
The advantage of the vacuum gluon solutions proposed in \cite{plb2018} is that non-Abelian dynamical 
gluon solutions possess Weyl symmetry and represent one-particle classical solutions corresponding to
one-particle quantum states after quantization. This phenomenon was observed first numerically \cite{plb2023}, 
and later it has
been proved by analytical methods in the Abelianized QCD \cite{plb2025}. This implies that one-particle Weyl symmetric 
gluon solutions possess an intrinsic color symmetry described by the singlet representation of  the Weyl group $W(SU(3))$.

Note, that standard QCD based on color group $SU(3)$ encounters a serious problem of  definition of 
elementary particles in the confinement phase due to a simple mathematical fact:
the group $SU(N)$ does not admit non-trivial irreducible singlet representations
with a color invariant one-dimensional subspace containing single particles. So that, the $SU(3)$
 fundamental quark triplet and gluon octet in adjoint representation can not be reduced to singlet one-particle multiplets.
In other words, the space of triplet quarks and octet gluons does not have non-trivial invariant one-dimensional
subspaces.
Therefore, the group $SU(3)$ admits only composite singlets made of quark and anti-quark,
or three antisymmetrized quarks, etc. Certainly, in the deconfinement phase the
single gluons and quarks may exist due to spontaneous color symmetry breaking, but in the color
confinement phase the color symmetry is preserved, and the mechanism of  spontaneous symmetry
breaking cannot be applied.

The second problem in introducing fundamental particles is related to the high non-linear structure of the
Yang-Mills equations which implies a hard problem of finding all one-particle quark and gluon
solutions satisfying complete exact equations of motion. This makes difficult to classify all
dynamical one-particle solutions and construct a Hilbert space of one-particle quantum states.
Fortunately, both problems related to existence of  color singlet non-pertubative
quark-gluon solutions can be resolved due to the rich group structure of $SU(3)$ Yang-Mills theory.
We will show that one can select a space of quark and gluon solutions which is invariant under the
action of a color subgroup of $SU(3)$, the Weyl group $W(SU(3))$, which provides a finite color
symmetry  of quarks and gluons after removal of all pure gauge gluon fields by means of gauge
fixing procedure and imposing a special solution ansatz.
It is really remarkable, that space of quark and gluon solutions contains invariant one-dimensional
vector subspaces consisting of invariant elements corresponding to quark and gluon singlets
belonging to standard singlet representations $\Gamma_1$ of the symmetric
group $S_3$ isomorphic to the Weyl group. This allows to classify all single quark and gluon
solutions \cite{plb2025}.

In the present paper we consider the color multiplet structure of gluon and quark fields on mass shell in $SU(N)$
Yang-Mills theories. We follow a non-perturbative approach considering solutions to the full
original non-linear Yang-Mills equations of motion.
Our approach is based on the idea that Weyl group symmetry is an inherent color symmetry of
gluons and quarks \cite{plb2023}.
 The Weyl group symmetry \cite{hamphreys} has been widely
considered in numerous physical models before: in various phenomenological 
Ginsburg-Landau type models of QCD 
\cite{KomaToki2000,KomaToki2001,TokiSuganuma2000}, in classification of color monopoles 
 \cite{cho1982,godd1977} and instantons  in $SU(3)$ Yang-Mills theory, and in studies of the 
 Weyl symmetric structure of the classical vacuum in QCD \cite{pak2012} etc.
Some applications of the Weyl group symmetry were considered in supersymmetric field models  \cite{supsymm}).
However, the close relationship of the Weyl group to color multiplets of fundamental particles, especially to singlet structure
of gluon and quark multiplets in $SU(N)$ Yang-Mills theory and QCD, have never been studied before.
\par
Our main purpose is to classify the quark and gluon solution subspaces which are invariant under
transformations of the Weyl group $W(SU(N))$ including generalization to $N\geq4$.
In a particular, we concentrate on the study of  invariant subspaces of one-particle quark and gluon
solutions, since the knowledge of  such
invariant single solutions provides strict definitions of fundamental particles. The plan of the
present paper is as follows. In Section II we consider the color multiplet structure of quark and
gluon fields  in $SU(3)$ Yang-Mills theory. The main results on multiplet
structure of $SU(3)$ QCD are obtained in \cite{plb2018,plb2023}, however, a strict
proof of the existence of a singlet gluon solution in a pure Yang-Mills theory was missed in 
previous studies.
We resolve this problem applying a strict mathematical definition of the Weyl group and show that
 singlet gluon represents an invariant element of a non-vector representation of the Weyl group.
 Our method implies an independent derivation of equations for the singlet and doublet constituent
 quarks  corresponding to one- and two-dimensional irreducible vector representations.
Section III is devoted to generalization of the obtained results in a pure $SU(3)$ QCD
to the case of the generalized QCD with the color group $SU(4)$.
Construction of standard Weyl group representations for the constituent quarks in Abelian projected
$SU(N)$ Yang-Mills theories with $N\geq 4$ is presented in Section IV. We demonstrate that
$SU(N)$ Yang-Mills theory manifests a universal structure of singlet quark-gluon solutions.
The last  section contains conclusions and discussion of possible applications of the presented results
 in QCD and hadron physics.

\section{Singlet gluon structure in $SU(3)$ Yang-Mills theory}
\label{sec:1}

Weyl multiplet structure of two- and three-particle gluon field solutions in $SU(3)$ Yang-Mills theory
(or pure QCD) is clearly determined by the vector representations $\Gamma_{2,3}$ of the 
Weyl group
\cite{plb2023,plb2025}. The most intriguing problem is the construction of a singlet one-particle gluon
solutions corresponding  to non-trivial irreducible one-dimensional  representations of the Weyl group.
Even in a simple case of a pure $SU(3)$ Yang-Mills theory there is a problem of the existence of a
Weyl singlet representation containing a singlet Weyl invariant gluon field. The problem is related to the
well known mathematical fact, that  in the standard basis of generators defined in terms of
Gell-Mann matrices, $T^a= 1/2 \lambda ^a$,
 two Abelian gluon fields $(A_\mu^3, A_\mu^8)$ form a vector space in the standard
 two-dimensional vector representation $\Gamma_2$, which is irreducible.
This implies that $\Gamma_2$ can not be reduced to a standard one-dimensional vector
 representation $\Gamma_1$. Therefore, it seems impossible to reduce the gluon
  doublet $(A_\mu^3, A_\mu^8)$ to a singlet, and, as a consequence, a single color 
  invariant gluon can not be defined as an
elementary particle, unless the color symmetry is broken.

In this section we revise the problem of the existence of a singlet gluon in a pure $SU(3)$
Yang-Mills theory.
The Lagrangian of $SU(3)$ Yang-Mills theory is defined in a standard manner
\bea
&&{\cal L}_{YM}=-\dfrac{1}{4} (F^{a\mu\nu} F^{a}_{\mu\nu}),
\eea
where $F_{\mu\nu}^a$ is the field strength tensor defined in terms
of the gauge vector potential $A_\mu^a$
$(\mu,\nu=0,1,2,3; a=1,2,...,8)$.
The Euler equations corresponding to the Yang-Mills Lagrangian ${\cal L}_{\rm YM}$ are given by a
system of coupled non-linear equations for the gauge field $A_\mu^a$
\bea
&&(D^\mu \vec F_{\mu\nu})^a=0, \label{eqA0}
\eea
where $D_\mu=\pro_\mu +\vA_\mu$ is a covariant derivative, and $\vA_\mu=A_\mu^a t^a$ is
a Lie algebra valued gauge connection, $t^a$ are generators in adjoint $SU(3)$ representation.
The commonly accepted multiplet structure of the gluon field is defined by the eight-dimensional
irreducible adjoint representation of the group $SU(3)$. Such a definition is valid only for gluon fields
off mass shell, i.e., for virtual gluons running in the internal lines and loops of Feynman diagrams, or,
the definition for gluon on shell is introduced in the framework of the perturbative QCD where gluon plane
wave solutions satisfy  truncated free equations of motion in the lowest order approximation.
The available experimental evidences of detection of eight gluons confirm
the number of free gluons in the scattering
processes with a large energy momentum transfer when such an approximation works well in the
perturbation theory, but free gluon solutions do not represent real physical particles. It is clear,
in the perturbative QCD, the conventional notions of gluons and quarks represent an artifact of using
approximate solutions to truncated equations of motion. We will consider color
multiplet structure of gluon solutions satisfying exact equations of motion.
In this respect, it is worth to notice, that gauge color symmetry is redundant in a sense that space
of physical dynamical gluon solutions is essentially smaller than a space of general gluon off-shell field 
configurations.
Due to this, it is suitable to remove all unphysical pure gauge gluon fields by imposing gauge
fixing conditions and/or applying some solution ansatz. This breaks down the initial $SU(3)$ color
symmetry of gluon fields to some subgroup of $SU(3)$. Therefore, the classification and color
attributes of gluon and quark solutions
become drastically different from the usual description based on color group $SU(3)$.
It is remarkable, after removing all pure gauge gluon fields the dynamical gluon solutions
still admit a color symmetry provided by the remaining finite color subgroup, the Weyl group,
which represents a real inherent color symmetry of gluons and quarks in old traditional sense.

A general Weyl-symmetric ansatz for dynamical $n$-particle axially-symmetric gluon solutions contains
all field components of the gauge potential $A_\mu^a$  located in color subspaces
corresponding to $I,U,V$-sectors defined explicitly in the Cartan-Weyl basis
 \cite{plb2023,flyvb,mpla2006}.
 We consider a reduced ansatz for one-particle gluon solutions in a pure $SU(3)$ QCD 
 with the folowing non-vanishing gluon fields 
 $A_\mu^a$ expressed in terms of four independent axially symmetric fields
 $K_\mu(t,r,\theta)$   \cite{plb2018}  (index values $(\mu=0,1,2,3)$ correspond to time
 and spherical space coordinates,  $x^\mu=(t,r,\theta,\varphi)$, respectively)
  \begin{align}
I:  &~~A_\varphi^1=q_1 K_3,~~~~~~  A_i^2=K_i,   \nn  \\
U: &~~A_\varphi^4=q_4 K_3, ~~~~~~ A_i^5=-K_i,    \nn \\
V: &~~A_\varphi^6=q_6 K_3, ~~~~~~ A_i^7 = K_i,   \nn \\
    &~~~~ A_\varphi^3=A_\varphi^8\equiv K_3, \label{A38}
\end{align}
where the values of parameters $q_1,q_4,q_6$ provide the consistency of the ansatz with all 
original Euler equations of motion (we changed slightly notations used in \cite{plb2018})
\bea
&&q_1=-2/\sqrt 3, ~~q_4= 1/\sqrt 3-1, ~~q_6= 1/\sqrt 3+1.
\eea
 Note, that there are well known no-go theorems \cite{deser76,coleman77} which forbid the existence 
 of localized stationary solutions  with a finite energy in the Yang-Mills theory. 
Our ansatz (3) defines spherical wave type non-Abelian solutions with a finite energy density
and infinite total energy in the whole space. The Abelian spherical wave type solutions were used 
in the old MIT bag models \cite{MIT1,MIT2} for description of propagating gluons inside the bag.
Certainly, the solutions can be quantized and produce quantum one-particle states, 
however, the problem of localizing such particles or quantum states in free empty space remains.
One of possible solutions to this problem is described in \cite{plb2023}, where it has been shown, that a single gluon
can be localized inside hadron due to interaction with the vacuum gluon condensate,
which leads to formation of the bound states corresponding to the lightest glueballs  (\cite{plb2023}, sections 5,6).
Note, that additional interaction terms appearing  in the quantum effective action due to quantum corrections 
break the scale invariance in the quantum theory. This allows to  remove restrictions imposed by no-go theorems.\\


Let us return to the properties of the ansatz (\ref{A38}).
The  $I,U,V$-type field variables of the Abelian gluon fields $A_\varphi^3, A_\varphi^8$
are defined with using roots $r^p_\alpha$ ($p=I,U,V$) and index values $\alpha=(3,8$) 
correspond to Cartan subalgebra generators $(T^3, T^8)$
\bea
&&A_\varphi^p=\sum_\alpha A_\varphi^\alpha r^p_\alpha=A_\varphi^3 r^p_3+A_\varphi^8 r^p_8, 
\label{AIUV} 
\eea
or, in the vector form, the root vectors are defined as follows
\bea
&& \vec r^I=(~1,~0), \nn \\
&& \vec r^U=(-\dfrac{1}{2},\dfrac{\sqrt 3}{2}), \nn \\
&& \vec r^V=(-\dfrac{1}{2},-\dfrac{\sqrt 3}{2}). \label{rootvec}
\eea
The ansatz (\ref{A38}) is symmetric with respect to the Weyl group which acts
by permutations on $I,U,V$-type gluon fields.
The Weyl multiplet structure determined by the ansatz is the following.
The Weyl group action of the triplet gluon fields $(A^2_i, -A_i^5, A_i^7)$ is defined by the standard
three-dimensional vector representation $\Gamma_3$ of the symmetric group $S_3$ (the sign minus
 in front of $A_i^5$ appears from the commonly accepted definition of the Gell-Mann matrix
 $\lambda^5$, so, the consistence of the ansatz with the Weyl symmetry and equations of motion
  requires an  opposite sign of the corresponding field component $A_i^5$).
  The representation $\Gamma_3$ is reducible, and, due to
   the constraints in (\ref{A38}), the gluon triplet $(A^2_i, -A_i^5, A_i^7)$ reduces
  to a one-dimensional singlet vector field $K_i$ ($i=0,1,2$)
\bea
(A^2_i, -A_i^5, A_i^7)=K_i(1,1,1),
\eea
where the constant color vector $(1,1,1)$ is a non-trivial singlet vector which is explicit invariant
under permutations of its components.
In a similar manner, the gluon triplet $(A_\mu^1, A_\mu^4,A_\mu^6)$ by simple 
renormalization of the gluon fields can be reduced to a singlet due to  ansatz (\ref{A38})
\bea
(\dfrac{1}{q_1} A_\varphi^1, \dfrac{1}{q_4} A_\varphi^4, \dfrac{1}{q_6} A_\varphi^6)=
                   K_3 (1,1,1),
\eea
under the condition that field $K_3$ is a Weyl singlet.
 
 The remaining two Abelian gluon fields $A_\varphi^3, A_\varphi^8$
 form a triplet made of Weyl field variables $A_\varphi^p$ which satisfy a constraint
\bea
A_\varphi^I+ A_\varphi^U+A_\varphi^V= 0. \label{Ap0}
\eea
So,  three fields $(A_\varphi^I, A_\varphi^U,A_\varphi^V)$ (or two independent Abelian fields
 $A_\varphi^{3,8}$, equivalently) belong to the standard two-dimensional vector representation 
 $\Gamma_2$. 
 Obviously, the last constraint in (\ref{A38}) breaks the Weyl symmetric structure of the vector 
 representation  $\Gamma_2$.  Therefore, the representation $\Gamma_2$ can not be reduced 
 to one-dimensional  representation $\Gamma_1$, and the field $K_3$ can not be a singlet, 
 causing the breaking of the Weyl symmetry  of the whole ansatz. 
 This leads to a puzzle: it seems that ansatz (\ref{A38}) can not define
  Weyl invariant singlet particles, since two Abelian fields $A_\varphi^{3,8}$ form the irreducible 
  representation $\Gamma_2$ which can not be reduced to the singlet representation $\Gamma_1$.
  On the other hand, the classical Lagrangian is explicit symmetric under the Weyl group permutations,
  and a careful analysis confirms that numerical solutions produced  by the ansatz
  represent one-particle solutions, each of them describes only one independent propagating magnetic
  polarization gluon mode  \cite{plb2018} (the electric type polarization mode is defined by a
  different dual electric ansatz \cite{plb2023}). We will perform a thorough consideration of the
  singlet Weyl group representations and resolve this puzzling issue.

  We start  with a strict  mathematical definition of the Weyl group and derive the
  action of the Weyl group on $SU(3)$ Lie algebra and on the space of gluon fields to determine
  the multiplet structure of gluon fields $A_\mu^a$.
  The Weyl group $W(G, T)$ for a given group $G=SU(3)$ and a torus $T$,
corresponding to the Abelian subroup of $SU(3)$, is defined as follows
\bea
W(G, T)=N(G, T)/Z(G, T),
\eea
where $N(G,T)$ is a normalizer of $T$ in $G$ and $Z(G, T)$ is a centralizer of $T$ in $G$.
In the most important case of the maximal Abelian subgroup $U_3(1)\times U_8(1)$
 the corresponding torus $T$ is maximal. With this, one has a relationship for
 the centralizer, $Z(G, T)=T$, and the Weyl group is defined simply by
\bea
W(G, T) = N(G, T)/T.
\eea
By definition, the normalizer is defined as follows
\bea
N(G,T)=\{g\in G: gtg^{-1}\in T ~\rm{for~all}~ t\in T\}. \label{Norm}
\eea
So that, we can find straightforward all group elements of $SU(3)$ which leaves the maximal torus
invariant under the adjoint group action in (\ref{Norm}).
As a final result, we have obtained the following six elements of the Weyl group
\bea
&L_1=\begin{pmatrix}
   0&1& 0\\
  1&0&0\\
   0&0&1\\
 \end{pmatrix},
 &L_2=\begin{pmatrix}
   1&0& 0\\
  0&0&1\\
   0&1&0\\
 \end{pmatrix},
 ~L_3=\begin{pmatrix}
   0&0& 1\\
  0&1&0\\
   1&0&0\\
 \end{pmatrix}, \nn\\
  &L_4=\begin{pmatrix}
   0&1& 0\\
  0&0&1\\
   1&0&0\\
 \end{pmatrix},
  &L_5=\begin{pmatrix}
   0&0& 1\\
  1&0&0\\
   0&1&0\\
 \end{pmatrix},
  ~L_6={\mathbb E}=\begin{pmatrix}
   1&0& 0\\
  0&1&0\\
   0&0&1\\
 \end{pmatrix}, \label{L456}
  \eea
where the last element ${\bbE}$ is a neutral element.
One can find the following properties of matrices $L_i$:
\bea
&&L_{1,2,3}^2=\bbE, \nn \\
&&L_{4,5}^3=\bbE,   \nn \\
&&L_4 L_5=L_5 L_4=\bbE.
\eea
 One can verify that elements $L_i$ $(i=1,2,...6)$ form a finite group of the sixth order, and
 each group element $L_i$ represents a permutation matrix which permutes components
 of a three vector $\vec V=(V_1,V_2,V_3)$ in a three-dimensional Euclidean space $\bbR^3$.
 Consider first the adjoint action of the Weyl group matrices $L_k$ on Cartan generators $(T^3,T^8)$
 and on corresponding Weyl $I,U,V$-type generators $T^p=\sum_{\alpha} T^\alpha \, r^p_\alpha$.
 Applying adjoint action of matrices $L_k$ on generators $T^\alpha$ ($\alpha=3,8$),
 \bea
 Ad_{L_k} T^\alpha= L_k T^\alpha L_k^\dagger,
\eea
  one obtains the following transformations
 of the Weyl triplet $T^p=(T^I, T^U, T^V)$
 \bea
 &&Ad_{L_1}(T^I, T^U, T^V)=(-T^I,-T^V,-T^U), \nn \\
 &&Ad_{L_2}(T^I, T^U, T^V)=(-T^V,-T^U,-T^I), \nn \\
 &&Ad_{L_3}(T^I, T^U, T^V)=(-T^U,-T^I,-T^V), \nn \\
 &&Ad_{L_4}(T^I, T^U, T^V)=(T^V,~T^I,~T^U), \nn \\
 &&Ad_{L_5}(T^I, T^U, T^V)=(T^U,~T^V,~T^I), \label{Tptrans}
 \eea
 where transformations of the triplet $T^p$ under the action of first three matrices $L_{1,2,3}$
 produce reflections of generators $(T^I,T^U,T^V)$ about the hyperplanes
 orthogonal  to each vector $T^p$, and transformations corresponding to matrices
 $L_{4,5}$ represent two kinds of cyclic permutations in opposite directions. So that, the
 transformations (\ref{Tptrans}) represent the symmetry group of the equilateral triangle formed
 by the ends of symmetric root vectors (\ref{rootvec}). This is precisely the original definition of the 
 Weyl group $W(SU(3),T)$ as a symmetry group of the root system of ${\mathfrak g}(SU(3))$.

   To find explicit realization of the Weyl group acting on the space of  Abelian gluon fields
   $A_\mu^{3,8}$ we  consider a Lie algebra valued Abelian gauge potential $\vA_\mu$
   decomposed in the constrained generator basis  $T^p$
   \bea
\vA_\mu=A_\mu^3 T^3+A_\mu^8 T^8 =\dfrac{2}{3}\big(A_\mu^I T^I+A_\mu^U T^U
                                                                    +A_\mu^V T^V\big), \label{Adecomp}
 \eea
 where Weyl field components $A_\mu^p \in \Gamma_2$ are expressed in terms of two Abelian
  fields $A_\mu^{3,8}$, (\ref{AIUV}). Since the vector $\vA_\mu$ is invariant and independent on a 
  choice of the generator basis, one can easily find Weyl group transformations of the fields
 $A_\mu^p$ and $A_\mu^{3,8}$ from the transformations rules (\ref{Tptrans}).
One can verify that Weyl transformations of $A_\mu^p$ are defined by the same equations
(\ref{Tptrans}) with the replacements $T^p\rightarrow A_\mu^p$. With this,  using relationship 
(\ref{Adecomp}) between two decompositions in $T^{3,8}$ and $T^p$ bases,
one can easily find explicit Weyl transformations of the fields $A_\mu^{3,8}$ defined by six
linear $(2 \times 2)$-matrices $\tL_k$ corresponding to matrices $L_k$ induced on the
two-dimensional vector space $(A_\mu^3, A_\mu^8) $.  For instance, matrices $\tL_1, \tL_2$,
 induced from the matrices $L_1, L_2$ have the following form
  \bea
  &\tL_1 =\begin{pmatrix}
   -1&~0\\
  0&~1\\
 \end{pmatrix},
  &\tL_2=\begin{pmatrix}
   1/2&~\sqrt 3/2\\
  \sqrt 3/2&~-1/2\\
 \end{pmatrix}.
  \eea
  So that, the Abelian fields $(A_\mu^3, A_\mu^8)$ form a doublet in the
  two-dimensional vector representation  $\Gamma_2$ of the Weyl group.
  The constraint $A_\mu^3=A_\mu^8$ is not invariant under two-dimensional Weyl transformations
  given by matrices  $\tL_k$, this implies that one can not reduce representation $\Gamma_2$ to
  one-dimensional vector representation $\Gamma_1$ in agreement with the general theory of
  vector representations of the symmetric group $S_3$.
  One can verify, that imposing constraint $A_\mu^I= A_\mu^U=A_\mu^V$ results in a trivial 
  singlet $\{ A_\mu^p\}  \equiv \{0,0,0\}$ as well. Nevertheless, we show that one can define a 
  non-trivial Weyl singlet gluon field outside the framework of vector representation theory.

 Consider three $I,U,V$-type two-component vectors in a two-dimensional plane $(X_3,X_8)$,
 each component represents $I,U,V$-projection of the vector potential $A_\mu^{3,8}$ with using
 the roots $r^p_{3,8}$
 \bea
 \vA_\mu^p=(A_\mu^3 r^p_3, A_\mu^8 r^p_8), \label{Ar}
 \eea
 or, in the explicit form
 \bea
 &&\vA^I_\mu=(A_\mu^3,~ 0), \nn \\
&& \vA^U_\mu=(-\dfrac{1}{2}A_\mu^3,~\dfrac{\sqrt 3}{2} A_\mu^8), \nn \\
&& \vA^V_\mu=(-\dfrac{1}{2}A_\mu^3,~-\dfrac{\sqrt 3}{2} A_\mu^8). \label{vecAp}
\eea
Consider the following constraint
\bea
A_\mu^3=A_\mu^8\equiv A_\mu^0. \label{constr1}
\eea
Under this constraint the vectors $\vA_\mu^p$, (\ref{Ar}), become proportional to
the root vectors, (\ref{rootvec}), as follows
 \bea
 \vA_\mu^p=A_\mu^0 \vec r^p.
 \eea
So that, vectors  $\vA_\mu^p$ possess the same symmetry as the system of three
symmetric root vectors $r^p_\alpha$. Remind, that original definition of
the Weyl group is introduced precisely as the symmetry group of the root system, which is isomorphic
to the symmetry group of the equilateral triangle.
Symmetry groups of regular geometric objects in three-dimensional Euclidean space give examples
of non-vector representations. The symmetry groups have some fixed points or lines
which represent invariant elements of the symmetric systems \cite{Weyl1952}. In our case of $SU(3)$
Yang-Mills theory the symmetry group of the system of vectors
$\vA_\mu^p$ admits a group of inner automorphisms which leave the field $A_\mu^0$ (\ref{constr1})
 invariant. Therefore, the invariant element $A_\mu^0$ represents a Weyl symmetric singlet gluon
 field in the non-vector representation of the Weyl group. In a case of unequal Abelian fields
 $(A_\mu^3, A_\mu^8)$,  the ends of vectors $\vA_\mu^p$, (\ref{vecAp}), make up the vertices of
 a non-equilateral triangle which in general does not have a symmetry group.
 However, the fields  $A_\mu^3, A_\mu^8$ still form an invariant two-dimensional vector space
 for the standard vector representation $\Gamma_2$. We conclude, for the equaled Abelian fields,
 (\ref{constr1}),  the vector system $\vA_\mu^p$ has the highest symmetry which is reflected by
 the existence of a non-Abelian symmetry group of the equilateral triangle formed by vector fields  
 $\vA_\mu^p$, and the vector system $\vA_\mu^p$ gains an invariant field element $A_\mu^0$
 which defines a basis vector in the invariant one-dimensional vector space.
 With this,  the vector field
 $A_\mu^0$ is naturally identified with the Weyl singlet gluon field in the non-vector representation
 of the Weyl group.

 Now, let us consider the role of equations of motion in the multiplet structure of gluon fields.
 The constraint (\ref{constr1}) has been obtained from the non-Abelian group structure of
 $SU(3)$, and it appears within the non-Abelian ansatz (\ref{A38}) which defines equations
 for one-particle solutions in a pure Yang-Mills theory. So, the constraint (\ref{constr1}) defines
 a singlet  gluon solution in a special case of equations of motion for Abelian gluon fields in a
 pure gluodynamics.  If singlet gluon solutions satisfy to different equations of motion, for example,
  to equations of motion in the presence of quarks, then the gluon multiplet structure will be changed.
   We will demonstrate, that in the presence of quarks one has two types of independent 
   singlet gluon fields.\\

       First we extend the adjoint action of the matrices $L_k$ to all generators of the
  Lie algebra of $SU(3)$.
  The adjoint action of matrices $L_k$ realize all possible permutations of generators
  $(T^1, T^4, T^6)$ and respective gluon fields ($A_\mu^1,A_\mu^4,A_\mu^6)$ forming
  an invariant vector subspace in three-dimensional Weyl representation $\Gamma_3$.
  Let us consider the following two generators
  \bea
  H_3=T^2 -T^5+ T^7, \label{H3su3}\\
  H_8=T^1+ T^4+ T^6. \label{H8su3}
  \eea
  The generators $(H_3,H_8)$ commute to each other, and they correspond to generators $T^3, T^8$
  defined in terms of Gell-Mann matrices. One can verify, that with a proper normalization the generators
  $H_3,H_8$ acquire the same eigenvalues as the standard generators $T^3, T^8$, but different
  eigenvectors.
  We show that generators $(H_3,H_8)$ form a Cartan subalgebra of ${\mathfrak g} (su(3))$ which is
  consistent with Weyl multiplet structure of quarks and gluons.
 The generator $H_8$ is invariant under the adjoint action of all Weyl matrices $L_k$,
 \bea
&&  Ad_{L_k} (H_8)=L_k H_8 L_k^\dagger= H_8, ~~{\rm for~all~} k,
\eea
 and the generator $H_3$ transforms as follows
  \bea
&& Ad_{L_k}(H_3)=-H_3,~~~k=1,2,3, \nn\\
&& Ad_{L_k}(H_3)=+H_3,~~~k=4,5,6.
  \eea
  This implies that $H_8$ represents an invariant element of Lie algebra ${\mathfrak g} (su(3))$ under
  all Weyl transformations,
  and the corresponding Abelian gluon field $A_\mu^8$ is a singlet in one-dimensional standard vector
representation
  $\Gamma_1$ of the Weyl group.
The generator $H_3$ is not strictly a singlet under transformations of matrices $L_{1,2,3}$,
but it becomes a singlet if we take into account that Weyl symmeric Lagrangian is invariant
under the reflections $A_\mu^a \rightarrow -A_\mu^a$
due to the mutual cancellation of all cubic gluon self-interaction terms after applying the
ansatz (\ref{A38}) \cite{plb2018}. In addition, the Yang-Mills Lagrangian written in the Cartan Weyl basis
is invariant under permutations of gluon fields from $I,U,V$ sectors, including permutations
of $I,U,V$-type gluon fields in the triplets $(A_\mu^1, A_\mu^4, A_\mu^9)$ and
$(A_\mu^2,A_\mu^5, A_\mu^7) $ corresponding to generators
 $(T^1,T^4,T^9)$ and $(T^2,-T^5, T^7)$, respectively. It is clear, that
both triplets can be reduced to two singlets by imposing the following constraints
\bea
&&A_\mu^1=A_\mu^4=A_\mu^9\equiv\tC_\mu^8, \\
&&A_\mu^2=A_\mu^5=A_\mu^7\equiv\tC_\mu^3.
\eea
With this,  one can introduce a new Weyl symmetric Abelian projection provided by the Cartan subalgebra
generators $(H_3, H_8)$
\bea
\vA_\mu=\tC_\mu^3 H_3+\tC_\mu^8 H_8,
\eea
 which implies the following Euler equations of motion
 \bea
&&(D_\mu F^{\mu\nu})^a= -\dfrac{g}{2} \bar \hPsi \gamma^\nu \lambda^a \hPsi, \\
&&\Big[i \gamma^\mu (\pro_\mu -\dfrac{ig}{2} \tC_\mu^3 H_3
                                   -\dfrac{ig}{2} \tC_\mu^8 H_8)-m\Big]\hPsi=0. \label{Euler3}
\eea
The Cartan generators $H_3, H_8$ have three common color eigenvectors
\bea
&&\hv^0=(1,1,1), \nn \\
&&\hv^\pm=\Big  (-\dfrac{1}{2}\pm i \dfrac{\sqrt 3}{2},
~-\dfrac{1}{2}\mp i \dfrac{\sqrt 3}{2}, ~ 1~\Big).\label{hv}
\eea
The most important feature of this new Cartan subalgebra basis is that generator $H_8$ has
a singlet color eigenvector $\hv^0\in \Gamma_1$ with a non-zero eigenvalue, whereas the
generator $H_3$ has eigenvector $\hv^0$ representing a zero mode. These properties
 of $H_3, H_8$ provide decomposition of the gluon representations $\Gamma_3$ into a
 sum of two irreducible gluon representations
 \bea
\Gamma_3=\Gamma_2\oplus \Gamma_1.
\eea
The singlet vector $\hv^0$ forms a basis in one-dimensional invariant vector space of the representation
 $\Gamma_1$, and two vectors $\hv^\pm$ form a basis in a two-dimensional invariant color space of the
 representation $\Gamma_2$.
These properties imply the decomposition of the quark triplet $\hPsi$ into a sum
of two irreducible Weyl multiplets, a singlet quark $\psi^0 \in \Gamma_1$ and a quark doublet
$(\psi^1, \psi^2) \in \Gamma_2$
\bea
&&\hPsi= \psi^0 \hv^0+\psi^1 \hv^++\psi^2 \hv^-. \label{Psidec3}
\eea
With this, one can define two solution ansatzes, for the singlet and for the doublet constituent quarks,
by the following reductions:\\
 the ansatz for the singlet quark $\psi^0$:
 \bea
 \hPsi=\psi^0 \hv^0, ~~~\psi^{1,2}=0; \label{singans}
 \eea
 the ansatz for the quark doublet $(\psi^1, \psi^2)$:
 \bea
 \hPsi=\psi^1 \hv^++\psi^2 \hv^- ~~~ \psi^0=0. \label{doubleans}
\eea
Substitution of the singlet ansatz (\ref{singans}) into Euler equations (\ref{Euler3})
leads to a simple system of equations
\bea
&&\pro_\mu F^{\mu\nu}[\tC^3]=0, \label{eqC3} \\
&&\pro_\mu F^{\mu\nu}[\tC^8]=-g \bar \psi^0 \gamma^\nu \psi^0,  \label{eqC8} \\
&&\big(i \gamma^\mu \pro_\mu -m + g \gamma^\mu \tC^{8\mu}\big) \psi^0=0, \label{psys2}
\eea
where the first equation describes a free gluon field $\tC_\mu^3$ which is completely decoupled,
and equations (\ref{eqC8}, \ref{psys2}) form a system of equations
for the constituent singlet quark $\psi^0$ dressed in singlet gluon field $\tC_\mu^8$.
Substitution of the second ansatz (\ref{doubleans}) into Euler equations equations  (\ref{Euler3})
 implies a system of
coupled equations for two constituent quarks $\psi^{1,2}$ and two dressing gluons $\tC_\mu^{3,8}$
belonging to irreducible representations $\Gamma_2$.

The equations (\ref{eqC8}, \ref{psys2}) for the singlet constituent quark have been obtained first in 
a recent paper \cite{plb2025} using a different approach based on constructing a full complex
Cartan-Weyl basis consistent with the Weyl multiplet quark-gluon structure. Such an approach allows
to establish one-to-one correspondence of gluon fields defined in the new full Cartan-Weyl basis
and gluon fields in the original formulation of QCD in the standard generator basis expressed in
terms of Gell-Mann matrices. In a particular, it has been established \cite{plb2025} that Abelian 
singlet gluon field
$\tC_\mu^8$ in standard vector representation $\Gamma_1$ is related to the Abelian singlet
gluon field $A_\mu^8$,  (\ref{constr1}), representing a singlet gluon in the non-vector
symmetry group representation, by a simple relationship
\bea
\tC_\mu^8=\sqrt 3 A_\mu^8,
\eea
where the number factor ``$\sqrt 3$'' reflects the presence of three gluon fields in the triplet
$(A_\mu^{1,4,6})$ before its reduction to one singlet gluon $\tC_\mu^8$. This number factor
results in a correct triple numerical coefficient in front of the one-loop correction to the effective action of
$SU(3)$ QCD calculated in past \cite{flyvb,mpla2006}.

Note that our conclusion on the existence of a Weyl singlet gluon field in the pure Yang-Mills theory
 is indirectly confirmed by the vacuum structure of the one-loop quantum effective potential
which has a unique absolute vacuum when the constant Abelian field strengths $H_{\mu\nu}^3$ 
and $H_{\mu\nu}^8$
are equaled by module \cite{plb2023}.

We conclude, even though the vector representation $\Gamma_2$ is irreducible,
the constraint (\ref{constr1}) implies that
vectors $\vA_\mu^p$ possess the symmetry group of the equilateral triangle which
realizes a non-vector representation of the Weyl group, and the gluon field  $A_\mu^0$, as an invariant 
element in the non-vector symmetry representation,
forms a singlet basis vector in the Weyl invariant one-dimensional space.
In the next section we will generalize the constraint  (\ref{constr1}) to the case of $SU(4)$
Cartan subalgebra and construct a Weyl singlet ansatz in a pure $SU(4)$ Yang-Mills theory.
     The existence of a non-trivial singlet and doublet structure of quarks and gluons is a
remarkable manifestation of the Weyl group symmetry, which represents a small discrete
color symmetry. However, the Weyl group inherits some principal features of the full color
group $SU(3)$ which are clearly manifested on mass shell.
One should note, that existence of a singlet gluon representation of the Weyl group implies
 the existence of a non-degenerate color invariant QCD vacuum,
as it has been supposed in old 1980s by Wilson, Kogut and Susskind \cite{wilson1974,kogut1975}.

Another interesting observation is that Zweig in his original unpublished work \cite{Zweig} 
considers the quark triplet as a sum of one singlet quark and one isospin quark doublet.


\section{Weyl singlet gluon solutions in $SU(4)$ Yang-Mills theory}

The  results presented in the previous section can be generalized straightforward to a case of a
$SU(N)$ Yang-Mills theory.
The classical Lagrangian of $SU(N)$ Yang-Mills theory is given in a standard form
\bea
&&\cL_{YM}=-\dfrac{1}{4} F^{a\mu\nu}F_{\mu\nu}^a. \label{LYM}
\eea
($a=1,2,...N^2-1$). In this section we consider Weyl multiplet structure in a pure
$SU(4)$ Yang-Mills theory without matter fields.
We use the generator basis $T^a=1/2 \lambda^a$ $(a=1,2,...15)$ in terms of $SU(4)$
Gell-Mann matrices.
 The roots are defined in the Cartan-Weyl basis containing three generators of the Cartan subalgebra ,
 $T^\beta$ $(\beta=3,8,15)$,  and six complex off-diagonal generators $T^p_\pm$
($p=1,2,...,6$)  spanning the coset space $SU(4)/U_3(1)\times U_8(1)\times U_{15}(1)$
\bea
&T^1_\pm=\dfrac{1}{2}(T^1\pm i T^2),~~&T^2_\pm=\dfrac{1}{2}(T^4\pm i T^5),\nn \\
&T^3_\pm=\dfrac{1}{2}(T^6\pm i T^7),~~ &T^4_\pm=\dfrac{1}{2}(T^9\pm i T^{10}),\nn \\
&T^5_\pm=\dfrac{1}{2}(T^{11}\pm i T^{12}),~~ &T^6_\pm
                                               =\dfrac{1}{2}(T^{13}\pm i T^{14}). \label{Tpm}
\eea
It is useful to pass to the Cartan-Weyl generator basis and introduce complex field
variables $W_\mu^p$ corresponding to off-diagonal generators $T^p_\pm$
\bea
&W_\mu^1=\dfrac{1}{\sqrt 2}(A^1_\mu+i A_\mu^2),
~~~~~&W_\mu^4=\dfrac{1}{\sqrt 2}(A^9_\mu+i A_\mu^{10}),\nn\\
&W_\mu^2=\dfrac{1}{\sqrt 2}(A^4_\mu+i A_\mu^5),
~~~~~&W_\mu^5=\dfrac{1}{\sqrt 2}(A^{11}_\mu+i A_\mu^{12}),\nn\\
&W_\mu^3=\dfrac{1}{\sqrt 2}(A^6_\mu+i A_\mu^7),~~~~~&
W_\mu^6=\dfrac{1}{\sqrt 2}(A^{13}_\mu+i A_\mu^{14}).  \label{complexW}
\eea
The roots can be defined as eigenvalues of three generators of the Cartan subalgebra,
 $T^\beta$ $(\beta=3,8,15)$, acting on the six generators $T^p_\pm$ by commutator
\bea
&& [ T^\beta, T^p_\pm]=\pm r^p_\beta T^p_\pm, \label{CWcom}
\eea
where six roots $ r^p_\beta$ corresponding to the positive sign in (\ref{CWcom}) read
 \bea
&\bfr^1=(1,0,0),~~~~~
                      ~~~~~~&\bfr^4=(\dfrac{1}{2}, \dfrac{1}{2 \sqrt 3}, \dfrac{2}{\sqrt 6}), \nn \\
&\bfr^2=(-\dfrac{1}{2} ,\dfrac{\sqrt 3}{2},0),~~
                   ~~~&\bfr^5=(-\dfrac{1}{2}, \dfrac{1}{2 \sqrt 3}, \dfrac{2}{\sqrt 6}),\nn \\
&\bfr^3=(-\dfrac{1}{2},-\dfrac{\sqrt 3}{2}, 0), ~
~~& \bfr^6=(0,-\dfrac{1}{\sqrt 3}, \dfrac{2}{\sqrt 6}). \label{roots5}
\eea
The remaining six roots corresponding to the negative sign in (\ref{CWcom}) complete a full system 
of twelve roots $\pm \bfr^p$.

In the Cartan-Weyl basis the $SU(4)$ Yang-Mills Lagrangian ${\cal L}_{YM}$  can be written in explicit
Weyl symmetric form
\bea
{\cal L}_{Weyl}&=&\sum_{p=1,...,6} \Big\{ -\dfrac{1}{8} (F^p_{\mu\nu})^2
+ \dfrac{1}{2}\big| D^p_{\mu} W_\nu^p-
 D^p_{\nu} W_\mu^p \big|^2
 -ig F_{\mu\nu}^p W^{*p}_{~\mu} W_\nu^p+ ...\Big\}, \label{Lweylsym}
 \eea
 where it is sufficient to use Weyl field variables of the field strength, $F^p_{\mu\nu}$,
 and  gauge potentials $A^p_\mu, W_\mu^p$, defined by root projections with using 
 only six roots $r^p_\beta$
 \bea
 &&F^p_{\mu\nu}=\pro_\mu A^p_\nu-\pro_\nu A^p_\mu,  \\
&&A^p_\mu=A^\beta_\mu r^p_\beta,  \label{Fp}
 \eea
and  $U(1)$ covariant derivatives $D^p_\mu$ is defined with the Abelian gluon field $A^p_\mu$
\bea
&&D^p_{\mu}W^p_\nu=(\pro_\mu -ig A^p_\mu)W_\nu^p.
\eea
The  quartic interaction terms in the Lagrangian ${\cal L}_{Weyl}$, (\ref{Lweylsym}),
are omitted since,  in a general, their explicit expressions depend on a chosen Weyl symmetric 
solution ansatz. 

  In a case of a pure $SU(4)$ Yang-Mills theory the Weyl group can be defined by a symmetry
 of the full root system, as it follows from the structure  of the Lie algebra $\mathfrak g(SU(4))$
 in the Cartan-Weyl basis $(T^\beta, T^p_\pm)$, (\ref{Tpm}). One can consider the roots 
 $\pm r^p_\beta$ as 
 twelve three-component root vectors $\pm \vec r^p=\pm (r^p_3, r^p_8, r^p_{15})$ in the 
 Cartan hyperplane spanned by generators $T^{3,8,15}$. With this, the ends of the twelve root 
 vectors $\pm \vec r^p$ form the vertices of a regular cuboctahedron,
 which has a high symmetry group containing the symmetric group $S_{12}$ which acts by 
 permutations of the vertices.

 However, the Lagrangian $\cL_{Weyl}$, (\ref{Lweylsym}),
  is explicit invariant under permutations of six $p$-type components of gluon fields
 $A_\mu^p, W_\mu^p$, which form the symmetric group $S_6$. Note, that Weyl group of $SU(N)$
  is not a symmetry group of roots for $N>3$. The standard definition of the Weyl group can be given
  as a symmetry of the Weyl chambers which is consistent with the definition of the Weyl group
  as a symmetry group of weights isomorphic to group $S_N$ \cite{hamphreys}. 
  Note that symmetry group $S_6$ of the Lagrangian $\cL_{Weyl}$
  is larger than finite color subgroup $S_4$ of $SU(4)$. Nevertheless, the presence of the symmetry  $S_6$
  simplifies the construction of the Weyl symmetric ansatz for singlet one-particle gluon solutions.
  
  The Weyl multiplet structure of gluon fields is determined by the following structure of the 
Cartan-Weyl basis or by the form of the Weyl symmetric Lagrangian $\cL_{Weyl}$ , (\ref{Lweylsym}).
One has one invariant subspace corresponding to
the Abelian fields $A_\mu^{3,8,15}$ forming three-dimensional space of the representation 
$\Gamma_3$, and one complex invariant subspace of complex gluon fields $W_\mu^p$ forming
the standard six-dimensional representation $\Gamma_6$ (or two real
invariant subspaces containing real off-diagonal gluon fields
$A_\mu^{1,4,6,9,11,13}$ and $A_\mu^{2,5,7,10,12,14}$ respectively).
All these representations are reducible and one can extract from them a singlet one-dimensional  
representation $\Gamma_1$ by imposing a proper ansatz for singlet gluon solutions.

We constrain our consideration with regular stationary non-Abelian solutions of color magnetic type,
and generalize the Weyl symmetric $SU(3)$ ansatz for singlet gluon solutions obtained in
 \cite{plb2018} to the case of $SU(4)$ Yang-Mills theory.
The magnetic type dynamical solutions are described by the vector gauge potential with the
azimuthal field component $A_\varphi$ in spherical space coordinates $(r,\theta, \varphi)$.
We introduce three Abelian fields of
magnetic type $A_\varphi^{3,8,15}$  and put
magnetic fields $A_\varphi^a$ into the color
space spanned by generators $T^{1,4,6,9,11,13}$ , while electric type fields $A_i^a$
 (index values $i=(0,1,2)$ correspond to space indices $(t,r,\theta)$) are placed
into the color space spanned by generators $T^{2,5,7,10,12,14}$. We need the electric
type potentials to provide the non-linear quartic Yang-Mills type interaction.

A general Weyl symmetric ansatz for many particle solutions contains three Abelian magnetic fields
$A_\varphi^{3,8,15}$ in the Cartan space, additional six magnetic fields
$A_\varphi^{1,4,6,9,11,13}$ and eighteen electric fields  $A_i^{2,5,7,10,12,14}$
corresponding to off-diagonal generators
\bea
&A_i^2=K_i,~~&A_i^5=Q_i,~~A_i^7=S_i,  \nn \\
&A_i^{10}=P_i,~~&A_i^{12}=T_i,~~A_i^{14}=R_i,  \nn \\
&A_\varphi^1=K_4,~~&A_\varphi^4=Q_4,~~A_\varphi^6=S_4,  \nn \\
&A_i^9=P_4,~~&A_i^{11}=T_4,~~A_i^{13}=R_4,\nn \\
&A_\varphi^3=K_3,~~&A_\varphi^8=K_8,~~A_\varphi^{15}=K_{15}. \label{SU4DHN}
\eea
 One can verify that ansatz (\ref{SU4DHN})
is consistent with all original $SU(4)$ Yang-Mills equations of motion, and the corresponding
Lagrangian is
symmetric under permutations of fields located in three invariant subspaces spanned by three
generator sets:
(i) Cartan subalgebra generators $T^{3,8,15}$; (ii) real coset space generators
$T^{1,4,6, 9,11,13}$, and (iii) pure imaginary coset space generators $T^{2,5,7,10,12,14}$.
The gluon fields
 $(K_i, Q_i, S_i, P_i,T_i,R_i)$, corresponding to $p$-components of the
 complex gluon fields $W^p_\mu$, (\ref{complexW}), form three invariant subspaces of electric 
 gluon fields in standard Weyl group representation $\Gamma_6$. In a similar way, the fields 
 $(K_4,Q_4,S_4,P_4,T_4,S_4)$ form one invariant subspace of magnetic gluon fields in the 
 representation $\Gamma_6$. The Abelian fields 
 $A_\varphi^{3,8,15}\equiv (K_3,K_8,K_{15})$ are equivalent to the sextet of the Weyl gluon fields 
 $A^p_\varphi=A_\varphi^\beta r^p_\beta $ forming a three-dimensional invariant
 subspace in the Weyl group representation $\Gamma_3$
 induced from the six-dimensional representation $\Gamma_6$ by root projection. 
 
One can reduce the general ansatz and construct a minimal Weyl symmetric ansatz which defines
only one-particle non-Abelian solutions. We impose the following reduction constraints
\bea
&&Q_4=q_4 K_4,~~S_4=q_6 K_4, \label{KQS1}\\
&&P_4=q_9 K_4,~~T_4=q_{11} K_4,~~R_4=q_{13} K_4, \label{KQS2} \\
&&K_3=c_3K_4, ~~K_8=c_8 K_4,~~K_{15}=c_{15} K_4, \label{K3815} \\
&&Q_i=p_2 K_i,~~S_i=p_3 K_i, \label{PTR1}\\
&&P_i=p_4 K_i,~~T_i=p_5 K_i,~~R_i=p_6 K_i, \label{PTR2}
\eea
where $c_3,c_8, c_{15}, p_k,q_k$ are free real number parameters, and
parameters $q_1, p_1$ can be absorbed by fields $K_4, K_i$ respectively, so we set $q_1=p_1=1$
up to normalization of the fields $K_4, K_i$. 

  The constraints (\ref{KQS1}-\ref{PTR2}) induce singlet structure of the invariant subspace in
 the coset spanned by $T^{1,4,6,9,11,13}$, and constraints (\ref{PTR1}, \ref{PTR2}) provide
 the singlet structure of the fields in the coset space spanned by  $T^{2,5,7,10,12,14}$.
With this, the reduction constraints allow to express all fields in terms of four independent fields
$(K_i, K_4)$ which form a four-dimensional vector field. One has to verify that  such an ansatz 
with the reduction constraints is consistent with all original equations of motion.
After substitution of the ansatz (\ref{SU4DHN}) and reduction constraints  (\ref{KQS1}-\ref{PTR2})
into the original equations of motion, we have obtained that all original Euler equations
 (including gauge fixing terms of Lorenz type \cite{plb2018}) reduce to a system of four independent
 second order partial differential equations and one quadratic constraint
 for four independent fields $K_\mu=(K_i, K_4)$ if all free parameters are expressed in terms of
 two parameters  $p_2, q_9$ as follows
\bea
&&p_3=-1-p_2,\nn \\
&&p_4=-1-p_2,\nn \\
&&p_5=2+p_2,\nn \\
&& p_6=-1, \label{p3456}
\eea
\bea
&&c_3=\dfrac{p_2(2+p_2)(-1+p_2-q_9)-2(1+q_9)}{2 p_2(1+p_2)},\nn \\
&&c_8=\dfrac{2(1+q_9)-p_2(2(1+q_9)+p_2(5+p_2+3q_9))}{2 \sqrt 3 p_2(1+p_2)},\nn\\
&&c_{15}=\dfrac{(2+p_2)(2(1+q_9)+p_2(3+p_2+3q_9))}{\sqrt   6 p_2 (1+p_2)}, \label{c3815}
\eea
\bea
&&q_4=-\dfrac{(2+p_2)(1+p_2+q_9)}{2 (1+p_2)},\nn \\
&&q_6=-\dfrac{1}{p_2}(2+2p_2+2q_9+p_2q_9),\nn \\
&&q_{11}=\dfrac{(2+p_2)(1+p_2+q_9)}{2(1+p_2)},\nn \\
&&q_{13}=\dfrac{1}{p_2}(2+p_2+2q_9).  \label{q461113}
\eea
Note, that three Abelian fields $K_3, K_8, K_{15}$, (\ref{K3815}), must be equaled to each other
\bea
K_3=K_8=K_{15},
\eea
 since only in this case the system of  twelve three-component Weyl vectors $\vA_\varphi^p$ 
 composed from the Abelian fields $(K_3, K_8, K_{15}) $ will be proportional to the full system of 
 root vectors
 \bea
 \vA_\varphi^p=(A_\varphi^3 r^p_3, ~A_\varphi^8 r^p_8, ~A_\varphi^{15} r^p_{15})
                                               =K_3 \vec r^p.
 \eea
 This implies that Abelian gluon field $K_3$ represents an invariant element of a non-vector
 symmetry group acting on the root system, i.e., a singlet with respect to the Weyl group $W(SU(4))$.
 With definitions (\ref{K3815})  one obtains an important constraint
 \bea
 c_3=c_8=c_{15},
 \eea
 which allows to solve three equations for $p_2,q_9,c_3$ in (\ref{c3815}) and obtain a unique regular
  solution in the analytical form
\bea
&&p_2=\dfrac{1}{2}(-6+3 \sqrt 2-2 \sqrt 3+\sqrt 6),\nn \\
&&q_9=\dfrac{1}{2}(152-57\sqrt 2+90 \sqrt 3-51 \sqrt 6),\nn \\
&&c_3=c_8=c_{15}=\dfrac{3}{23} (-12+16 \sqrt{2}-18 \sqrt{3}+\sqrt{6}).\nn \\
\eea
With this one has the following final expressions for the remaining parameters
\bea
&&q_4=\dfrac{1}{46}(-24+9 \sqrt 2+10\sqrt 3+25\sqrt 6), \nn \\
&&q_6=\dfrac{1}{46} (-40+153 \sqrt{2}-106 \sqrt{3}+11 \sqrt{6}), \nn \\
&&q_{11}= \dfrac{1}{46} (24-9 \sqrt{2}-10 \sqrt{3}-25 \sqrt{6}), \nn \\
&&q_{13}=\dfrac{1}{23}(-79-48 \sqrt{2}+8 \sqrt{3}+20 \sqrt{6}).
\eea
It is interesting to notice, that parameters $q_i, p_i$ satisfy the following identities
\bea
&&q_1+q_4+q_6+q_9+q_{11}+q_{13}=0,  \nn \\
&& p_1+p_2+p_3+p_4+p_5+p_6=0, \label{pident}
\eea
which imply reduction of the sextets ${\hat U}_4$ and ${\hat V}_i$ defined as follows
\bea
&&{\hat U}_4=(K_4, Q_4, S_4, P_4, T_4, R_4), \nn \\
&&{\hat V}_i=(K_i, Q_i, S_i, P_i, T_i, R_i).
\eea
It is clear, that ${\hat U}_4$ and ${\hat V}_i$
form four six-dimensional subspaces invariant under permutations of the group 
$S_6$ acting on component fields of ${\hat U}_4$ and ${\hat V}_i$
 corresponding to the initial gluon fields $A_\varphi^{1,4,6,9,11,13}$ and $A_i^{2,5,7,10,12,14}$.
 After proper renormalization by dividing these initial gluon fields by respective parameters
 $q^{1,4,6,9,11,13}$ and $p_{1,2,3,4,5,6}$ the sextets ${\hat U}_4, {\hat V}_i$ turn into
 gluon singlets
 \bea
 &&{\hat U}_4=K_4(1,1,1,1,1,1) \in \Gamma_1, \nn \\
 &&{\hat V}_i=K_i(1,1,1,1,1,1) \in (\Gamma_1)_i.
 \eea
 Note, that singlet ${\hat U}_4$ is consistent with the
singlet structure of the Abelian gluon fields $K_3, K_8, K_{15}$ representing an invariant element of the
gluon fields corresponding to the Cartan subalgebra $SU(4)$ due to constraints (\ref{KQS1}-\ref{PTR2}).

Finally, one can write down the Lagrangian of $SU(4)$ Yang-Mills theory
for one-particle Weyl symmetric gluon solutions in the following form
\bea
\cL^{\it Weyl\,\, sym.}(K)&=&\cL^{(2)}+\cL^{\it g.f.}+\cL^{(4)},  \\
\cL^{(2)}&=&-\dfrac{c_1}{2 r^4 \sin^2\theta}\big ( (\pro_\theta K_3)^2-r^2 (\pro_t K_3)^2+
                             r^2(\pro_r K_3)^2\big),\nn \\
&&+\dfrac{ c_2}{2} \Big ((\pro_t K_1- \pro_r K_0)^2+\dfrac{1}{r^2} (\pro_t K_2 -\pro_\theta K_0)^2
-\dfrac{1}{r^2}(\pro_\theta K_1-\pro_r K_2)^2 \Big ) \\
\cL_{\it g.f.}&=&-\dfrac{c_2}{2} \big (\pro_r K_1-\pro_t K_0+\dfrac{1}{r^2}\pro_\theta K_2\big)^2, \\
\cL^{(4)}&=&-\dfrac{c_1c_2}{8 r^2\sin^2\theta} K_3^2\big ( K_1^2-K_0^2+\dfrac{1}{r^2}K_2^2\big),
\eea
where 
\bea
&&c_1=\dfrac{2}{9}(307-192 \sqrt 2-160 \sqrt 3+112 \sqrt 6)=7.263... , \\
&&c_2=1+p_2^2+p_3^2+p_4^2+p_5^2+p_6^2 =4(14-9\sqrt 2+7\sqrt 3-5\sqrt 6)=4.596...\, .
\eea

By construction the Lagrangian $\cL^{\it Weyl\,\, sym.}(K)$ contains the  quadratic part of the
Yang-Mills Lagrangian, $\cL^{(2)}$, Lagrangian $\cL^{\it g.f.}$ with the gauge fixing terms, 
and Lagrangian $\cL^{(4)}$ with quartic  interaction terms. All cubic interaction terms are mutually 
canceled due to the Weyl symmetry.
All Lagrangian parts, $\cL^{(2)}$, $\cL^{\it g.f.}$ and $\cL^{(4)}$ coincide with the corresponding 
parts of the
Weyl symmetric $SU(2)$ and $SU(3)$ pure Yang-Mills theories \cite{plb2018,plb2023} up to number 
coefficients
$c_1, c_2,$ which are determined uniquely by corresponding groups $SU(2)$ and $SU(3)$. So that,
we conjecture
that Weyl symmetric Lagrangian for one-particle singlet gluon solutions has a universal structure for 
the gauge group $SU(N)$
for arbitrary values of $N$. The origin of such universality comes from the structure of the Weyl symmetry
which provides the symmetry of $SU(2)$ subgroups from various Weyl $p$-sectors, making all of 
them equivalent. Due to this, 
the final Weyl symmetric Lagrangian depends on only one four vector field $K_\mu=(K_i, K_\varphi)$,
and uniqueness of the form of the Lagrangian $\cL^{\it Weyl\,\, sym.}(K)$ implies
that Weyl symmetry provides a non-trivial embedding of the subgroup $SU(2)$ into the full group $SU(N)$ 
which preserves the highest color symmetry of singlet gluon solutions. 
 
We conclude, that imposing reduction constraints (51-55) reduces a general ansatz (50) 
to the ansatz for singlet one-particle gluon solutions defined in terms of four 
independent fields $K_\mu=(K_i, K_4)$.
The singlet ansatz is consistent with all original Euler equations and with the original Yang-Mills
Lagrangian, and it implies a system of four hyperbolic second order partial differential equations
which have the same qualitative form as the equations for singlet gluon solutions in $SU(2)$ and
$SU(3)$ Yang-Mills theories. We expect that such a qualitative structure of the Weyl symmetric
Lagrangian for singlet gluon solutions takes place in $SU(N)$ Yang-Mills theory for $N>4$
This can justify non-perturbative  methods in QCD based on models with color group $SU(N)$
in large $N$ approximation.\\

\section{Weyl symmetric Abelian projection of $SU(N)$ Yang-Mills theory with quarks}

It was conjectured by t' Hooft that in the confinement phase the vacuum structure
should be described by the Abelian projected
QCD \cite{thooft81}. It is supposed that at low energies in infra-red regime the Abelian
dominance effect plays an important role \cite{abeldom1,abeldom3,plb2023}, and
low energy quantum states in QCD are described in the leading order by the Abelianized QCD.
So that, one expects that one should try to find a singlet quark solution, if it exists, in the
Abelian projected QCD. However, it turns out that in ordinary Abelian projection
with the standard Cartan subalgebra basis $T^3, T^8$ the QCD equations do not reduce to
equations for non-zero Weyl symmetric singlet quark solutions belonging to one-dimensional representation
$\Gamma_1$ of the Weyl group. To construct equations for a singlet quark  one needs
to use a special Abelian projection which admits irreducible Weyl representations
$\Gamma_{1,2}$ for quarks and dressing gluons as well, as it has been found in a
case of $SU(3)$ QCD in Section II. We apply our method of constructing equations for the
singlet quarks to a general case of $SU(N)$ Yang-Mills
theory with quarks in $SU(N)$ fundamental representation.

We consider a full Lagrangian of $SU(N)$ Yang-Mills theory
 containing the quark Lagrangian $\cL_q$ with one quark multiplet in $SU(N)$ fundamental
 representation
\bea
&&{\cal L}={\cal L}_{YM}+{\cal L}_{ q}, \nn \\
&&{\cal L}_{q}= \bar \hPsi \Big[ i \gamma^\mu (\pro_\mu - i g A_\mu^a T^a)-m\Big]\hPsi, \label{Lq}
\eea
where $F_{\mu\nu}^a$  is the field strength tensor defined in terms
of the gauge vector potential $A_\mu^a$ in a standard manner,
 ${\cal L}_q$ is the Lagrangian of matter fields described by one fundamental quark multiplet
$\hPsi=(\psi_1, \psi_2,..., \psi_N)$.
The Euler equations of motion  corresponding to the full Yang-Mills Lagrangian represent
a system of non-linear coupled equations for the gauge fields $A_\mu^a$ and for the
fermion multiplet $\hPsi=(\psi^1, \psi^2,..., \psi^N)$
\bea
&&(D^\mu \vec F_{\mu\nu})^a=-\dfrac{g}{2}\bar \hPsi \gamma_\nu\lambda^a\hPsi,\label{eqA1}\\
&& \Big[ i \gamma^\mu (\pro_\mu-i \dfrac{g}{2} A_\mu^a\lambda^a)-m\Big]\hPsi=0.\label{eqPsi1}
\eea\\

{\bf 4.1 Weyl symmetric Abelian projection in\\
 $SU(4)$ Yang-Mills theory}\\

In a case of $SU(N)$ Yang-Mills theories with quark matter for values $N\geq 4$
it is suitable to introduce the Weyl group as a symmetry group of weights,
which appears naturally as a symmetric group $S_N$ acting by permutations of
$SU(N)$ quark fields in the Lagrangian $\cL_q$, (\ref{Lq}).
 In a case of $SU(4)$ Yang-Mills theory one has three Cartan subalgebra
 generators $T^\beta=1/2 \lambda^{3,8,15}$ which have four common color eigenvectors $\hu^i$
\bea
&&T^\beta \hu^i=w^{ \beta i} \hu^i,  \nn\\
&& \hu^1=(1,0,0,0),~~~\hu^2=(0,1,0,0), ~~~\hu^3=(0,0,1,0), ~~~ \hu^4=(0,0,0,1),
\eea
where eigenvalues $w^{\beta i}$ define four weight vectors $\bfw^i$ in a
 three-dimensional Cartan hyperplane spanned by $(T^3, T^8, T^{15})$
\bea
&&\bfw^1=(~\dfrac{1}{2},\dfrac{1}{2\sqrt 3},\dfrac{1}{2\sqrt 6}), \nn \\
&&\bfw^2=(-\dfrac{1}{2}, \dfrac{1}{2\sqrt 3},\dfrac{1}{2\sqrt 6}), \nn \\
&&\bfw^3= (~0,-\dfrac{1}{\sqrt 3},\dfrac{1}{2\sqrt 6}), \nn \\
&&\bfw^4=(~0,~0,~ -\dfrac{3}{2\sqrt 6}).
\eea
The weights $w^{\beta i}$ define color charges of quarks and coupling constant corresponding
to quark-gluon interaction terms.
The ends of the weight vectors make up the vertices of the regular tetrahedron which
has a symmetry group isomorphic to the symmetric group $S_4$ \cite{hamphreys}
 acting by permutations of the vertices of the tetrahedron. With this, the symmetry of weights
 defines naturally the standard four-dimensional vector representation $\Gamma_4$ realized by
  linear matrix transformations in the four-dimensional vector space. The representation
  $\Gamma_4$ corresponds to $SU(4)$ fundamental quark quadruplet.

With this,  the  quark Lagrangian $\cL_q$, (\ref{Lq}), can be written
 in an explicit factorized Weyl symmetric form
\bea
 \cL_q&=& \sum_i \bpsi^i \Big (i \gamma^\mu \pro_\mu -m +
 \dfrac{g}{2}
\gamma^\mu (A_\mu^i)_{(w)}\Big )\psi^i, \label{lagrqp} \\
&&(A^i_\mu)_w=A_\mu^\beta w^{i\beta},
\eea
where the Weyl symmetric $i$-field component $(A^i_\mu)_w$ is defined with using weights.

It is known, that root vectors can be defined in terms of the weights by the following relationships
\bea
\vec r^p=\vec w_i-\vec w_k, ~~~ i\neq k, \label{rootweight}
\eea
where index $p$ of the root vector is defined by a pair of weight indices, $p=(i,k)$. Totally one has
twelve roots and four weights. It is clear, that Weyl group $W(SU(4))$  acts on the root space
by the induced four-dimensional representation $\Gamma_4$ defined by the relationships
 (\ref{rootweight}). Note, that  induced representation $\Gamma_4$ acts on all twelve roots
 $\pm r^p_\beta$ by permutations from the subgroup of $S_{12}$. One can define action of the
  induced representation $\Gamma_4$ in the space of six roots $r^p_\beta$ as well, since
  reflections $r^p_\beta \rightarrow -r^p_\beta$, imply reflections of gluon fields corresponding
  to off-diagonal generators $T^p_\pm$, (\ref{Tpm}).
So that the equivalence relationship for roots with respect to reflections is justified by
the invariance of the Weyl symmetric Lagrangian under reflections of gluon fields in the color
subspaces spanned by generators $T^{1,4,6,9,11,13}$ and $T^{2,5,7,10,12,14}$.
Such an invariance of the Lagrangian under reflections of off-diagonal glon fields is caused
 by disappearance of cubic gluon interaction terms  after applying
 the Weyl symmetric ansatz (\ref{KQS1}-\ref{PTR2}).

Generalization of our results to $SU(4)$ Yang-Mills theory with fermions in fundamental representation
can be done straightforward along the lines of derivation of the equations for singlet constituent quark
 in $SU(3)$ QCD \cite{plb2025}.
A principal step in this direction is the construction of the Cartan subalgebra generators
which must be consistent with the multiplet structure of quarks under the transformations
of the Weyl group $W(SU(4))$.
The Cartan subalgebra of $\mathfrak g(SU(4))$ contains three commuting generators $H^\beta$
($\beta=3,8,15$).
The Cartan generator $H^{15}$ is defined in a similar way as in $SU(3)$ QCD
\bea
  H^{15}&=&T^1+T^4+T^6+T^9+T^{11}+T^{13}
 = \dfrac{1}{2}\begin{pmatrix}
   0&1&1&1\\
  1&0&1&1\\
   1&1&0&1\\
    1&1&1&0\\
 \end{pmatrix}.
\label{H15}
 \eea
 An important feature of the generator $H^{15}$ is that generator possesses a singlet color
 eigenvector  $\hu^0=(1,1,1,1)$ with a non-zero eigenvalue. Note, that a standard basis
 of generators  $T^a=1/2 \lambda^a$ defined in terms of known generalized Gell-Mann matrices
 $\lambda^a$ is not  consistent with Weyl multiplet structure of quarks, since none of
 generators $T^a$ admits a singlet color eigenvector with a non-zero eigenvalue. Due to this,
 the equations of motion in the standard Gell-Mann generator basis are inconsistent with the
 singlet quark structure and admit only unphysical free quark solutions.
 The remaining two generators of the Cartan superalgebra are
defined as follows
\bea
  H^{8}&=&T^1+T^4-2T^6-2T^9+T^{11}+T^{13}
 = \dfrac{1}{2}\begin{pmatrix}
   0&1&1&-2\\
  1&0&-2&1\\
   1&-2&0&1\\
    -2&1&1&0\\
 \end{pmatrix}, \label{H8su4}
\qquad\\
 H^3&=&T^2-T^5+T^{12}-T^{14}
 =\dfrac{1}{2}\begin{pmatrix}
 0&-i & i&0\\
  i &0&0&-i\\
   -i&0&0&i\\
  0&i&-i&0\\
 \end{pmatrix}.
\label{H3su4}
 \eea
 One can verify that matrices $H^3, H^8, H^{15}$ are analogs of corresponding Gell-Mann matrices
 $\lambda^3, \lambda^8, \lambda^{15}$, and after proper normalization the matrices
 $H^\beta$  have the same eigenvalues as matrices $\lambda^\beta$, respectively.
The matrices $H^\beta$  commute to each other and form the Cartan subalgebra of
 $\mathfrak g(SU(4))$ which defines an Abelian  projection of $SU(4)$ Yang-Mills theory
 to Abelian gauge theory with the following Abelian gauge vector potential
  \bea
 &&\vA_\mu = \tC_\mu^{15} H^{15}+\tC_\mu^{8} H^{8}+\tC_\mu^{3} H^{3}.
 \eea
 The Lagrangian of $SU(4)$ Abelian projected Yang-Mills theory with $SU(4)$ fundamental quarks
 takes a simple form
 \bea
 \cL_{\it Abel.}=-\dfrac{1}{4} \sum_\beta(F_{\mu\nu}[\tC^\beta])^2
              +\bar \hPsi \Big( i \gamma^\mu \pro_\mu -m)
           +g \sum_\beta \gamma^\mu \tC_\mu^\beta H^\beta \Big) \hPsi,\label{LAbel}
 \eea
 where $F_{\mu\nu}[\tC^\beta]=\pro_\mu\tC_\nu^\beta- \pro_\nu\tC_\mu^\beta$.

The generators $H^\beta$ have four common eigenvectors $\hu^k$ ($k=0,1,2,3$)
\bea
\hu^0&=&\dfrac{1}{2}(1,1,1,1),~~~\hu^1=\dfrac{1}{2}(1,-1,-1,1),~~~
\hu^2=\dfrac{1}{2}(-1,-i,i,1),~~~\hu^3=\dfrac{1}{2}(-1,i,-i,1),  \label{w0123}
\eea
where eigenvectors $\hu^k$ form an orthonormal basis, $(\hu^k)^\dagger \hu^n=\delta^{kn}$.
The vector  $\hu^0$ represents a singlet vector which is a basis vector in one-dimensional
vector space of the standard Weyl group representation $\Gamma_1$. Other three vectors
 $\hu^{1,2,3}$ form the vector basis
in a three-dimensional space of the second standard representation $\Gamma_3$ of $S_4$. 
So that, the Cartan subalgebra generators $H^\beta$ provides decomposition
of the four dimensional vector space into a product of two invariant subspaces corresponding
 to two irreducible Weyl group representations $\Gamma_1$  and $\Gamma_3$.
 Therefore, one can perform the following decomposition of the fundamental $SU(4)$ fermion quadruplet
 $\hPsi$ in the vector basis $\hu^k$,
\bea
\hPsi=\psi^0 \hu^0+\psi^1 \hu^1+\psi^2 \hu^2+\psi^3 \hu^3, \label{Psidec4}
\eea
which provides an explicit decomposition of the fermion multiplet $\hPsi$ into a sum of irreducible multiplets
containing one singlet quark $\psi^0$ from the representation $\Gamma_1$ and
one triplet $(\psi^1,\psi^2, \psi^3)$ from the representation $\Gamma_3$. Thus, the decomposition
(\ref{Psidec4}) implies the decomposition of the reducible representation $\Gamma_4$ into a sum of 
two irreducible representations
\bea
\Gamma_4= \Gamma_3\oplus\Gamma_1.
\eea
To obtain the Weyl multiplet structure of $SU(4)$ quarks on mass shell, we apply two ansatzes
for the singlet and triplet quark solutions.
The singlet ansatz is defined by the non-vanishing singlet term in the decomposition (\ref{Psidec4})
with setting other quark fields to zero
\bea
&&\hPsi=\psi^0 \hu^0,~~~~~ \psi^{1,2,3}=0. \label{singpsi}
\eea

Using the singlet ansatz (\ref{singpsi}) one can rewrite the full original equations of motion
(\ref{eqA1},\ref{eqPsi1}) in the following form
\bea
&&\pro^\mu F_{\mu\nu} [C^\beta]=-g\bar\hPsi \gamma_\nu H^\beta \hPsi,
                                                                                    \label{eqA2}\\
&& \Big( i \gamma^\mu \pro_\mu -m + g \sum_{\beta
             =3,8,15}C_\mu^\beta H^\beta \Big)\hPsi=0, \label{eqPsi2}
\eea
where $F_{\mu\nu} [C^\beta]$ is a field strength defined by the Abelian field $C_\mu^\beta$
corresponding to the Cartan generator $H^\beta$. Using the ansatz (\ref{singpsi}) and properties of
 generators $H^\beta$
one obtains a final system of equations for the singlet constituent quark $\psi^0$ dressed in a
singlet gluon $C_\mu^{15}$ and for two free Abelian gluon fields $C_\mu^{3,8}$ which decouple
from other fields
\bea
&&\pro^\mu F_{\mu\nu} [C^\beta] =0, ~~~{\rm for}~ \beta = 3,8,  \label{freeA} \\
&&\pro^\mu  F_{\mu\nu} [C^{15}] = -\dfrac{3}{2}g\bar \psi^0 \gamma_\nu \psi^0,\label{eqAfin} \\
&& \Big( i \gamma^\mu \pro_\mu -m + \dfrac{3}{2}g C_\mu^{15} \Big)\psi^0=0. \label{eqPsifin}
\eea

The ansatz for the quark triplet $(\psi^1, \psi^2, \psi^3)$ from the irreducible representation
$\Gamma_3$ is defined by the remaining part of the decomposition (\ref{Psidec4}) with the vanishing
 quark field $\psi^0$
\bea
&&\hPsi=\psi^1 \hu^1+\psi^2\hu^2+\psi^3 \hu^3, ~~~~~~\psi^0=0.\label{trippsi}
\eea
Substitution of this ansatz into Euler equations (\ref{eqA2}, \ref{eqPsi2}) produces a system of
coupled equations for three constituent quarks dressed in three Abelian gluon fields $\tC_\mu^\beta$.
\bea
&&\pro_\mu F^{\mu\nu}[\tC^{15}]=-g \sum_k \bpsi^k \gamma^\nu \psi^k,\nn \\
&&\pro_\mu F^{\mu\nu}[\tC^8]=4g\bpsi^1\gamma^\nu\psi^1-2g\bpsi^2\gamma^\nu\psi^2
-2g\bpsi^3\gamma^\nu\psi^3,\nn\\
&&\pro_\mu F^{\mu\nu}[\tC^3]=-2g \bpsi^2 \gamma^\nu \psi^2
                                                                  +2g \bpsi^3 \gamma^\nu \psi^3,\nn \\
&&\big(i\gamma^\mu\pro_\mu-m+g\gamma^\mu\tC^{15}_\mu
-4g\gamma^\mu\tC_\mu^8\big)\psi^1=0, \nn\\
&&\big(i\gamma^\mu\pro_\mu-m+g\gamma^\mu\tC^{15}_\mu
+2g\gamma^\mu\tC_\mu^8+2g\gamma^\mu\tC_\mu^3\big)\psi^2=0,\nn\\
&&\big(i\gamma^\mu\pro_\mu-m+g\gamma^\mu\tC^{15}_\mu
+2g\gamma^\mu\tC_\mu^8+2g\gamma^\mu\tC_\mu^3\big)\psi^3=0,\label{Gamma3eqs}
\eea
where all quark-gluon coupling constants are defined by eigenvalues of generators $H^\beta$.
The quark-gluon solutions satisfying the equations (\ref{Gamma3eqs}) belong to the second
standard irreducible representation $\Gamma_3$ of $S_4$.\\

{\bf 4.2 Abelian projection in $SU(N)$ ($N\geq 5$) Yang-Mills theory}\\

Cartan subalgebra of $SU(5)$ group consistent with Weyl singlet structure of quark multiplets
can be constructed in a similar manner as in the case of $SU(4)$ group.
We present final main results related to $SU(5)$ Weyl symmetric Abelian projection which provides the
decomposition of quark fundamental quintet into a sum of two irreducible Weyl multiplets.
 Four Cartan subalgebra generators $H^\gamma$ ($\gamma=3,8,15,24$) consistent with Weyl 
 multiplet structure of quarks and gluons are defined as follows
 \bea
  H^{24}
 &=& \dfrac{1}{2}\begin{pmatrix}
   0&1&1&1&1\\
  1&0&1&1&1\\
   1&1&0&1&1\\
   1&1&1&0&1\\
    1&1&1&1&0
 \end{pmatrix}, \label{h24}
\qquad\\
 H^{15}
 &=&\dfrac{1}{2}\begin{pmatrix}
 -3&2 & -3&2&2\\
  2&-3&-3&2&2\\
   -3&-3&12&-3&-3\\
  2&2&-3&-3&2\\
  2&2&-3&2&-3
 \end{pmatrix}.
\label{h15}
 \eea
 \bea
  H^8
 &=&\dfrac{1}{2} \begin{pmatrix}
   0&1&0&1&-2\\
  1&0&0&-2&1\\
   0&0&0&0&0\\
   1&-2&0&0&1\\
    -2&1&0&1&0
 \end{pmatrix}, \label{H8su5}
\qquad\\
 H^3
 &=&\dfrac{1}{2}\begin{pmatrix}
 0&-1 & 0&1&0\\
  -1&0&0&0&1\\
   0&0&0&0&0\\
  1&0&0&0&-1\\
  0&1&0&-1&0
 \end{pmatrix}.
\label{H3su5}
 \eea
The generators $H^\gamma$ have five common color eigenvectors
\bea
\hc^0&=&(1,1,1,1,1), \nn \\
\hc^1&=&(-1,1,0,-1,1), \nn \\
\hc^2&=&(-1,-1,0,1,1), \nn \\
\hc^3&=&(1,-1,0,-1,1), \nn \\
\hc^4&=&(1,1,-4,1,1). \label{u01234}
\eea
The color vector $\hc^0$ is a singlet vector of the one-dimensional representation $\Gamma_1$, and
color vectors $\hc^k$ $(k=1,2,3,4)$ form the basis in the vector space of the standard
four-dimensional irreducible representation $\Gamma_4$.
After proper normalization the generators $H^\gamma$ form a Cartan subalgebra of $SU(5)$
which is isomorphic to the Cartan subalgebra defined in
the basis of usual Abelian generators $T^\gamma =1/2 \lambda^\gamma$
defined in terms of $SU(5)$ Gell-Mann matrices. Both generator bases $H^\gamma$
and $T^\gamma$ have the same eigenvalues and root systems. So that, the Weyl group is
uniquely defined as the symmetry group of roots or weights.
A Weyl symmetric  Abelian projection is defined by the decomposition of the Abelian gauge vector
 potential in the generator basis $H^\gamma$
\bea
\vA_\mu=\tC^\gamma_\mu H^\gamma.
\eea
Using decomposition of $SU(5)$ fermion multiplet $\hPsi$  in the vector basis
$\hc^k$,
\bea
\hPsi=\psi^k \hc^k=\psi^0 \hc^0+\psi^1 \hc^1+\psi^2 \hc^2+\psi^3 \hc^3+\psi^4 \hc^4,
\eea
 one can define two ansatzes for quark-gluon
solutions from two standard Weyl group representations $\Gamma_1$ and $\Gamma_4$, in a full
analogy with the cases of $SU(3)$ and $SU(4)$ Yang-Mills theories.

Now it is clear, the generalization of the obtained results to a case of $SU(N)$ group for
 $N>5$ can be done straightforward.
One can verify that equations for the singlet quark from the singlet representation $\Gamma_1$
is unique, therefore one has only one type of a singlet constituent quark in $SU(N)$ Yang-Mills theory,
and remaining $N-1$ constituent quarks form a standard (N-1)-dimensional irreducible Weyl multiplet
belonging to representation $\Gamma_{N-1}$.

In the general case of $SU(N)$ Yang-Mills theory the ansatz for a singlet quark is defined by one 
non-vanishing quark field $\tpsi^0$, other quarks $\tpsi^k$ ($k=1,2,...,N-1$) are vanished and 
do not appear in the solution for the singlet constituent quark,
\bea
&&\hPsi=\tpsi^0 \hc^0. \label{ansSUN}
\eea
Substitution of the ansatz (\ref{ansSUN}) into equations of motion (\ref{eqA1}, \ref{eqPsi1})
results in the following system of equations (after simple rescaling of the fields $\tC^r_{\mu},\, \psi^0$)
\bea
&&\pro^\mu F_{\mu\nu}[\tC^i_{\mu}]=0,  \label{eqA2N} \\
&&\pro^\mu F_{\mu\nu}[\tC^r_{\mu}]=
                           -\dfrac{g (N-1)}{2}\bar \psi^0 \gamma_\nu \psi^0, \label{eqA1N} \\
&&\big(i \gamma^\mu \pro_\mu -m
                                  + \dfrac{g (N-1)}{2}\tC^r_{\mu}\big) \psi^0=0, \label{psys2N}
\eea
where $\tC^i_{\mu}$ ($i=1,2,...,r-1$) are $N-2$ free Abelian gluon fields  
corresponding to Cartan generators
 $H_i$ ($i\leq r-1$), ``$r$'' is the rank of the group $SU(N)$.
One has only one constituent singlet quark $\tpsi^0 \in \Gamma_1$ dressed in one singlet
gluon $\tC^r_\mu$. The free gluon fields  $\tC^i_{\mu}$ are completely decoupled and 
can be set to zero.

Other $N-1$ quarks $\psi^{k}$ ($k>1,..., N)$) form irreducible $(N-1)$-dimensional Weyl multiplet
$\Gamma_{N-1}$ which describes $(N-1)$ constituent quarks interacting to $N-1$ Abelian gluon
fields $\tC_\mu^i$.
Equations of motion for $N-1$ constituent quarks from representation $\Gamma_{N-1}$
can be obtained by substitution  of the following ansatz into the Euler equations 
(\ref{eqA1}, \ref{eqPsi1})
\bea
&&\psi^0=0, \nn \\
&&\hPsi=\sum_{k=1,2,...,r}\psi^k \hc^k.
\eea
The constituent quarks $\psi^k$ with dressing gluons $\tC^k$ $(k=1,2,...,N-1)$ form irreducible 
$(N-1)$-dimensional multiplet in the standard $(N-1)$-dimensional representation $\Gamma_{N-1}$ 
of the Weyl group.\\

 \section{Discussion}
 
 We have considered the multiplet structure of quarks and gluons on mass shell
 in $SU(N)$ Yang-Mills theory. In a case of pure $SU(3)$ Yang-Mills theory the existence of 
 a singlet gluon solution has been revised by applying strict mathematical methods of the representation
 theory of the Weyl group. It has been shown that singlet gluon solution to exact equations of motion
 represents an invariant element of a non-vector symmetry group
 representation of the Weyl group. 
 A natural question arises whether such Weyl singlet structure of gluon can be observed in the real QCD. 
 It is surprising, that a simple manifestation of the Weyl multiplet structure of gluon fields was found  long time ago
 in the structure of the one-loop quantum effective potential $V_{\it{eff}}$ of QCD calculated by Flyvbjerg 
  in 1980 \cite{flyvb}]. 
 The effective potential $V_{\it{eff}}$ is a gauge invariant physical quantity corresponding to the vacuum energy density which is a functional 
 depending on  two Abelian color magnetic gluon potentials $A_\varphi^3, A_\varphi^8$  representing the vacuum gluon 
 condensate functions, 
 which can be written in terms of $I,U,V$-type Weyl variables $A_\varphi^p$,  (5). In the constant field approximation 
 the effective potential  $V_{\it{eff}}$ represents a function of three scalar variables $F_p^2=(F_p^{\mu\nu})^2$
 corresponding to the field strengths $F_p^{\mu\nu}$ defined in terms of the gluon potentials $A_\varphi^p$. 
 The Weyl field variables 
 $F_p^2$ are gauge invariant and can be expressed in terms of three Casimir invariants $C_2, C_4, C_6$. In  \cite{flyvb,mpla2006}
it has been shown, that effective potential has six degenerate local minimums and one absolute minimum which corresponds
  to the deepest vacuum energy at some value of the vacuum gluon condensate. The six local minimums form the
   Weyl sextet corresponding to two-dimensional representation $\Gamma_2$ of the Weyl group, and the absolute minimum
   forms the Weyl singlet belonging to one-dimensional representation $\Gamma_1$. The Weyl sextet and singlet are described
   by different sets of the Casimir invariants $C_{2n}$. The details of the Weyl multiplet structure of the effective potential 
   are presented in (\cite{plb2025}, section 3).\\ 
\par
  The method based on using the non-vector representation of the Weyl group $W(SU(3))$ is generalized  
 to $SU(4)$ Yang-Milld theory in order to construct an explicit ansatz for the singlet gluon field.
 Generalization to a pure $SU(N)$  Yang-Mills theory is straightforward and implies a universal 
 structure of the final Weyl symmetric Lagrangian expressed in terms of the four-dimensional vector
 gluon field $K_\mu$ which describes one non-Abelian singlet gluon solution of magnetic type.
 
 The multiplet structure of quark Weyl multiplets in $SU(N)$ Yang-Mills theory has been studied 
 by using the Abelian projection in a new Cartan-Weyl generator basis 
 which is consistent with the Weyl multiplet structure of quarks and gluons.
 This allows to derive systems of equations for the constituent quarks dressed in gluon fields.
 It has been demonstrated that there are two types of irreducible Weyl multiplets for the constituent 
 quarks;
 one singlet quark dressed in one singlet gluon in the standard one-dimensional representation 
 $\Gamma_1$, and one standard irreducible $(N-1)$-dimensional multiplet for the constituent quarks
 dressed in $(N-1)$ Abelian gluons.\\
 
  One should stress that dynamical singlet quark-gluon solutions to classical equations of motion  (38, 39)
represent physical observable particles only in the deconfinement phase. In the confinement phase
the quarks are localized inside hadrons, so classical solutions for the constituent quarks 
can not provide such localization. To obtain physical solutions for the confined singlet quarks 
 one should take into account the interaction of the quarks with the vacuum gluon and quark condensates.
  It has been shown in (\cite{plb2023}, section 5), that interaction
 of a single primary gluon with the vacuum gluon condensate leads to formation of localized bound states 
 corresponding to the lightest glueballs. The obtained analytical expression for the glueball spectrum is 
 consistent with the Regge theory.
  We expect that interaction of the constituent singlet quark to vacuum gluon condensate will produce
new hadrons like the hybrid states made of gluons and quarks. This issue will be considered in a separate paper.

\par
The presented non-trivial Weyl group multiplets
 of quarks and gluons admits the existence of singlet and doublet color quarks
 in the standard $SU(3)$ QCD. This implies that the old traditional quark model in QCD should be 
 improved taking into account the non-perturbative multiplet structure of quarks and gluons.
 Such a modified quark model will describe more detailed spectrum of known hadrons and
 exotic mesons.\\
 
         {\bf Data availability statement}\\

All data that support the findings of this study are included within the article.

\acknowledgements

One of authors (DGP) thanks Prof. I.V. Anikin for invitation to visit BLTP JINR and for
numerous inspiring discussions.
This  work is supported by Chinese Academy of Sciences  (PIFI Grant No. 2019VMA0035),
National Natural Science Foundation of China (Grant No. 11575254),
and by Japan Society for Promotion of Science (Grant No. L19559).

\vskip 1 cm

\appendix

{\bf APPENDIX}\\

{\bf Vector representations of the Weyl group }   \\

The Weyl group $W(SU(N))$ is defined as a symmetry group of weights which is isomorphic to the
 symmetric group $S_N$.
In most applications of the Weyl group symmetry in physical models it is suitable to use 
a well known theory of vector representations of the group $S_N$ which has three standard 
vector representations.
The N-dimensional standard vector representation $\Gamma_N$ of $S_N$ is defined by 
all possible
permutations of  the components of a given vector $\vec V=(V_1, V_2,...V_N)$ in N-dimensional
vector space $\bbR^N$. 
So that, the representation space of $\Gamma_N$ is the $N$-dimensional vector space,
and Weyl group transformations can be realized by linear matrix transformations in the vector space.
The representation $\Gamma_N$ is reducible, and it can be decomposed into a
 direct sum of two standard irreducible representations
\bea
\Gamma_N=\Gamma_{N-1} \oplus \Gamma_1.
\eea
The $(N-1)$-dimensional representation $\Gamma_{N-1}$
 is defined by inducing the representation $\Gamma_N$ to $(N-1)$-dimensional hyperplane
 defined by a constraint imposed on the vector components
\bea
V_1+V_2+....+V_N=0. \label{plane}
\eea

A standard one-dimensional vector representation $\Gamma_1$ is defined by imposing
the following constraint on the vector  components
\bea
V_1=V_2=...=V_N\equiv V. \label{vec1}
\eea
It is evident that all vectors with equaled components are invariant under permutations of $S_N$.
Effectively, the standard representation $\Gamma_1$ is equivalent to the trivial representation containing
only one neutral element of the group. However, one should stress, that representation
$\Gamma_1$ is non-trivial since it includes a representation space ${\cal V}^1$  
containing all vectors with equaled components which form a non-trivial one-dimensional vector space 
described by a straight line passing through two points $P_0=(0,0,...,0)$ and $P_1=(1,1,...,1)$. 
Such a space consists of vectors all of which remain unchanged under matrix transformations of 
the group $S_N$, i.e., each vector in ${\cal V}^1$
represents a non-trivial invariant element under the action of the group $S_N$ isomorphic to the Weyl 
group $W(SU(N))$.
The existence of non-trivial invariant elements under group automorphisms
is an important property  of the group structure of $SU(N)$ \cite{Weyl1952}.
Remind, that group $SU(N)$ does not admit non-trivial one-dimensional representations.
The only vector, either in fundamental or adjoint representation or other ones, which is invariant under
all $SU(N)$ transformation, is a trivial zero vector, which does not form one-dimensional vector space.
This implies in a particular case of Yang-Mills theory with the gauge group $SU(3)$
 that it is impossible to introduce the concept of a single quark
 precisely due to the absence of such invariant one-dimensional
subspaces in the three-dimensional color space of quarks unless the color symmetry is broken.
So that, in the conventional QCD it is only possible to define composite color singlets.
\end{document}